\documentclass[a4paper,11pt]{article}
\usepackage{caption}
\usepackage{jcappub}

\usepackage{lineno}
\usepackage{xcolor}
\usepackage{multirow}
\usepackage{array}
\usepackage{graphics}
\usepackage{hhline}
\usepackage{makecell}
\usepackage{arydshln}
\usepackage{subcaption}
\usepackage{graphicx}
\usepackage{hyperref}
\usepackage{soul}
\usepackage[capitalise,nameinlink,noabbrev]{cleveref}
\usepackage{orcidlink}

\newcommand{\likelihood}{\mathcal{L}(\boldsymbol{d}|\boldsymbol{\theta})}

\newcommand{\deltachisquaredtheta}{\Delta \chi^2(\theta_i)}
\newcommand{\posterior}{P(\boldsymbol{\theta}|\boldsymbol{d})}
\newcommand{\prior}{p(\boldsymbol{\theta})}

\makeatletter
\input{aas_macros.sty}
\let\jnl@style=\relax
\makeatother

\title{Frequentist Cosmological Constraints from Full-Shape Clustering Measurements in DESI DR1}

\author[1,2]{J.~Morawetz\orcidlink{0009-0008-1034-1665},}
\author[1,2]{H.~Zhang\orcidlink{0000-0001-6847-5254},}
\author[1,2,3]{M.~Bonici\orcidlink{0000-0002-8430-126X},}
\author[1,2,3]{W.~J.~Percival\orcidlink{0000-0002-0644-5727},}
\author[1,2]{A.~Crespi\orcidlink{0009-0009-9437-0232},}
\author[4]{J.~Aguilar,}
\author[5]{S.~Ahlen\orcidlink{0000-0001-6098-7247},}
\author[6,7]{D.~Bianchi\orcidlink{0000-0001-9712-0006},}
\author[8]{D.~Brooks,}
\author[9,10]{F.~J.~Castander\orcidlink{0000-0001-7316-4573},}
\author[4]{T.~Claybaugh,}
\author[11]{S.~Cole\orcidlink{0000-0002-5954-7903},}
\author[4]{A.~Cuceu\orcidlink{0000-0002-2169-0595},}
\author[12]{A.~de la Macorra\orcidlink{0000-0002-1769-1640},}
\author[13]{A.~de~Mattia\orcidlink{0000-0003-0920-2947},}
\author[14,15]{Biprateep~Dey\orcidlink{0000-0002-5665-7912},}
\author[8]{P.~Doel,}
\author[4,16]{S.~Ferraro\orcidlink{0000-0003-4992-7854},}
\author[17]{A.~Font-Ribera\orcidlink{0000-0002-3033-7312},}
\author[18,19]{J.~E.~Forero-Romero\orcidlink{0000-0002-2890-3725},}
\author[9,10,20]{E.~Gaztañaga\orcidlink{0000-0001-9632-0815},}
\author[4,21]{S.~Gontcho A Gontcho\orcidlink{0000-0003-3142-233X},}
\author[22]{G.~Gutierrez,}
\author[23]{C.~Hahn\orcidlink{0000-0003-1197-0902},}
\author[24,25,26]{K.~Honscheid\orcidlink{0000-0002-6550-2023},}
\author[27,28]{D.~Huterer\orcidlink{0000-0001-6558-0112},}
\author[29]{M.~Ishak\orcidlink{0000-0002-6024-466X},}
\author[30]{R.~Joyce\orcidlink{0000-0003-0201-5241},}
\author[31]{R.~Kehoe,}
\author[32]{D.~Kirkby\orcidlink{0000-0002-8828-5463},}
\author[4]{T.~Kisner\orcidlink{0000-0003-3510-7134},}
\author[8]{O.~Lahav,}
\author[4]{A.~Lambert,}
\author[4]{M.~Landriau\orcidlink{0000-0003-1838-8528},}
\author[33]{L.~Le~Guillou\orcidlink{0000-0001-7178-8868},}
\author[17,34]{M.~Manera\orcidlink{0000-0003-4962-8934},}
\author[17,35]{R.~Miquel,}
\author[36]{E.~Mueller,}
\author[20]{S.~Nadathur\orcidlink{0000-0001-9070-3102},}
\author[15]{J.~ A.~Newman\orcidlink{0000-0001-8684-2222},}
\author[37,38]{G.~Niz\orcidlink{0000-0002-1544-8946},}
\author[4,13]{N.~Palanque-Delabrouille\orcidlink{0000-0003-3188-784X},}
\author[39]{F.~Prada\orcidlink{0000-0001-7145-8674},}
\author[40]{I.~P\'erez-R\`afols\orcidlink{0000-0001-6979-0125},}
\author[41]{G.~Rossi,}
\author[42,43,44]{L.~Samushia\orcidlink{0000-0002-1609-5687},}
\author[45]{E.~Sanchez\orcidlink{0000-0002-9646-8198},}
\author[4]{D.~Schlegel,}
\author[27,28]{M.~Schubnell,}
\author[4]{J.~Silber\orcidlink{0000-0002-3461-0320},}
\author[30]{D.~Sprayberry,}
\author[28]{G.~Tarl\'{e}\orcidlink{0000-0003-1704-0781},}
\author[30]{B.~A.~Weaver,}
\author[33]{P.~Zarrouk\orcidlink{0000-0002-7305-9578},}
\author[4]{R.~Zhou\orcidlink{0000-0001-5381-4372},}
\author[46]{H.~Zou\orcidlink{0000-0002-6684-3997},}

\affiliation{Affiliations are in \cref{sec:affiliations}}
\emailAdd{jgmorawe@uwaterloo.ca}

\abstract{We present a frequentist analysis of clustering measurements from Data Release 1 of the Dark Energy Spectroscopic Instrument (DESI) using the standard profile likelihood method. While Bayesian inferences for effective field theory models of galaxy clustering can be highly sensitive to prior choices for extended cosmological models, frequentist inferences are not susceptible to such effects. We compare frequentist and Bayesian constraints for the parameter set $\{\sigma_8, H_0, \Omega_{\rm{m}}, w_0, w_a\}$ using the full-shape power spectrum multipoles, post-reconstruction baryon acoustic oscillation (BAO) measurements, and external datasets from the CMB and type Ia supernovae measurements. The frequentist confidence intervals are significantly shifted relative to the Bayesian credible intervals for the $w_0w_a$CDM model, unless supernovae data are included. When DESI full-shape and BAO data are fit jointly, we obtain the following $1\sigma$ frequentist confidence intervals for $\Lambda$CDM ($w_0w_a$CDM): $\sigma_8 = 0.863^{+0.048}_{-0.040} , \ H_0 = 68.96^{+0.81}_{-0.80} \ \rm{km \ s^{-1}Mpc^{-1}} , \ \Omega_{\rm{m}} = 0.3034\pm0.0110$ ($\sigma_8 = 0.782^{+0.060}_{-0.036} , \ H_0 = 63.7^{+4.2}_{-2.0} \ \rm{km \ s^{-1}Mpc^{-1}} , \ \Omega_{\rm{m}} = 0.378^{+0.024}_{-0.047}$ , \ $w_0 = -0.16^{+0.10}_{-0.50}$ , \ $w_a = -3.0^{+1.7}_{}$), corresponding to 0.8$\sigma$, 0.3$\sigma$, 0.7$\sigma$ (2.1$\sigma$, 4.1$\sigma$, 6.5$\sigma$, 6.3$\sigma$, 6.6$\sigma$) shifts between the maximum likelihood estimate and the Bayesian posterior mean for $\Lambda$CDM ($w_0w_a$CDM) respectively.}

\begin{document}
\maketitle
\flushbottom

\section{Introduction}
\label{sec:intro}

Large-scale structure (LSS) observations are a pillar of modern cosmology, providing key insights about the universe's origin and evolution. This wealth of information is often derived from the three-dimensional spatial clustering of galaxies, with their distribution in comoving space inferred from redshifts and angular positions. Encoded within clustering measurements is the baryon acoustic oscillation (BAO) feature, which serves as a robust standard ruler for measuring cosmic expansion. Additionally, modeling the full shape of the clustering signal encodes details about the growth of structure and the amplitude and shape of the primordial power spectrum.

Previous-generation surveys, including the Baryon Oscillation Spectroscopic Survey (BOSS)~\cite{Dawson_2013} and its extension eBOSS~\cite{Dawson_2016}, have significantly advanced our understanding of cosmic expansion and structure growth. New-generation surveys are now realizing a significant increase in the quantity of available data. Among these projects is the Dark Energy Spectroscopic Instrument (DESI)~\cite{DESI_Collaboration_2016, DESI_Collaboration_2022, Miller_2024, Poppett_2024, Guy_2023, Schlafly_2023}, the first Stage-IV galaxy survey in operation. It is conducting a five-year spectroscopic program over 14,200 square degrees of the sky, measuring redshifts for more than 50 million galaxies and quasars. DESI uses five distinct tracers – the Bright Galaxy Sample (BGS), Luminous Red Galaxies (LRG), Emission Line Galaxies (ELG), Quasars (QSO), and the Lyman-$\alpha$ (Ly$\alpha$) forest – to cover a broad redshift range of $0 < z < 4$. The first data release (DESI DR1)~\cite{DESI_Collaboration_2025} included spectra from one year of observations, and the resulting science~\cite{DESI_BAO_VI_2024, DESI_FS_VII_2024} has provided interesting constraints on cosmological models.

The existing analyses conducted by DESI~\cite{DESI_BAO_VI_2024, DESI_FS_VII_2024, DESI_DR2_BAO} have largely focused on inference within the Bayesian statistical framework. In this approach, the posterior distribution – the probability density for the parameters conditioned on the observed data – is obtained by multiplying the likelihood function by the prior distribution and normalizing the result according to Bayes' theorem. The likelihood describes the probability of observing the data given fixed parameters, while the prior is a probability density over parameter space encapsulating existing knowledge or assumptions. Uniform priors are often considered ``non-informative" on the grounds that they do not explicitly favor specific parameter values. However, this characterization is misleading, since priors are probability densities; a prior that has a uniform distribution in one parameterization generally transforms (via the Jacobian) to a distribution that is non-uniform under a nonlinear change of variables~\cite{Trotta_2008}. Priors with uniform distributions in different parameter bases assign different probabilities to the same set of physical models. In the absence of a clear theoretical or physical motivation for selecting a preferred parameterization, the coordinate dependence associated with adopting uniform priors is arbitrary and can exert undue influence on the resulting parameter inference. The Jeffreys prior~\cite{Jeffreys_1946} is a commonly used alternative because it defines a coordinate-independent rule for prior selection, based on Fisher information, that ensures equivalent prior weightings when independently constructed in different parameterizations.

Previous studies such as~\cite{Simon_2023, Holm_2023a} have separated the impact of priors into two categories: prior weight effects (PWE) and prior volume ``projection" effects (PVE). PWE arise when the prior is not well aligned with the likelihood, leading to shifts of the maximum a posteriori (MAP) parameters relative to the maximum likelihood estimate (MLE) parameters in a given parameter basis. PVE occur when large regions of prior-supported but low-likelihood nuisance parameter space dominate the projection of the posterior onto lower-dimensional parameter subspaces, causing the resulting marginal posterior to shift significantly away from the MAP. However, this distinction is itself parameterization-dependent because, unlike the MLE, the MAP is not invariant under reparameterization. Thus, a shift between the marginal posterior and the MLE reflects a combination of PWE and PVE that is challenging to disentangle and non-trivial to interpret. Consequently, we will refer to both contributions collectively as prior effects. At the same time, priors are necessary and can be beneficial when they are physically well motivated (e.g. HOD-informed priors \cite{Ivanov_2024, Zhang_2025a, Akitsu_2024, Zhang_2025b}).

These prior effects have been studied in previous full-shape (FS) analyses of galaxy clustering data~\cite{Holm_2023a, Simon_2023} and have been under scrutiny – particularly for extended cosmological models – in recent DESI analyses~\cite{DESI_FS_V_2024, DESI_FS_VII_2024}. In FS analyses, the complete clustering signal is fitted using perturbation theory (PT)-based models~\cite{Bernardeau_2002} and augmented by an effective field theory (EFT) approach to capture small-scale effects~\cite{Baumann_2012, Carrasco_2012, Porto_2014, Perko_2016, Lewandowski_2017, Chen_2020, DAmico_2021, Philcox_2022, Ivanov_2022}. However, this approach also introduces a large number of nuisance parameters that must be marginalized over. Many of these parameters are only weakly constrained by the data and are highly degenerate with cosmological parameters, making the inference susceptible to the prior effects discussed above~\cite{Carrilho_2023, DAmico_2022, Holm_2023a}. Several approaches have been studied to mitigate these issues within the Bayesian framework, including introducing physically motivated priors based on the galaxy-halo connection~\cite{Ivanov_2024, Zhang_2025a, Akitsu_2024, Zhang_2025b}, developing different integration measures~\cite{Reeves_2025}, employing non-linear reparameterizations to decorrelate the nuisance parameters and the cosmological parameters~\cite{Paradiso_2024, Tsedrik_2025}, and applying the Jeffreys prior~\cite{Hadzhiyska_2023, Gsponer_2024, Bonici_Jeffreys_2025}. Another option is to derive cosmological constraints within the frequentist framework~\cite{Holm_2023a, Herold_2025b}.

The Bayesian interpretation of uncertainty characterizes knowledge in terms of degrees of belief, constructing credible intervals from the posterior probability distribution, which combines prior information with observed data. In contrast, the frequentist interpretation regards uncertainty as arising from the long-term frequency of events, treating model parameters as fixed but unknown quantities. Frequentist methods construct confidence intervals such that, under repeated experimentation, a specified proportion of these intervals contain the true parameter value, a property known as coverage. This approach is inherently parameterization-invariant and relies solely on observed data without incorporating prior beliefs, thereby sidestepping prior effects altogether. Neither Bayesian nor frequentist methodologies are objectively superior; rather, they address fundamentally different inferential questions. Employing both paradigms concurrently, as done in this analysis, can yield a deeper and more nuanced interpretation of scientific results.

Frequentist methods for parameter estimation and interval construction have a rigorous foundation in classical statistical inference, prominently formalized by Neyman~\cite{Neyman_1937}, who defined confidence intervals using the likelihood at the observed data across a range of models. A further improvement to the methodology was introduced by Feldman \& Cousins~\cite{Feldman_1998}, which resolved issues inherent in naive confidence interval constructions, particularly in boundary cases or parameter regions near physical limits. Feldman-Cousins intervals systematically handle both two-sided and one-sided intervals, providing a unified procedure that maintains correct coverage probability. Another powerful technique is the profile likelihood method, which approximates the Neyman and Feldman-Cousins approaches in certain limits but is more tractable in practice when many nuisance parameters and multidimensional data are present. This method constructs intervals by profiling the likelihood function over nuisance parameters and directly relating intervals to variations in the likelihood. The profile likelihood method benefits significantly from Wilks' theorem~\cite{Wilks_1938}, which establishes that, under regularity conditions, the distribution of twice the log-likelihood ratio asymptotically follows a chi-squared distribution. This connection simplifies interval estimation and hypothesis testing in frequentist analyses, enabling the robust and straightforward construction of confidence intervals. A recent cosmology-focused review of frequentist methods is given in~\cite{Herold_2025a}.

Our paper is organized as follows. In Section~\ref{sec:data}, we discuss the data and modeling used in our analysis. In Section~\ref{sec:method}, we discuss the methodology applied for the frequentist and Bayesian approaches. In Section~\ref{sec:results}, we present our main findings including a direct comparison of frequentist and Bayesian constraints. Finally, in Section~\ref{sec:summary}, we conclude by summarizing our findings and discussing plans for future work.

\section{Data \& Modeling}
\label{sec:data}

To ensure consistency with the baseline Bayesian FS analysis~\cite{DESI_FS_V_2024, DESI_FS_VII_2024}, we adopt the same data and models used in that work. This includes inference using four distinct datasets: FS power spectrum multipoles and post-reconstruction BAO measurements (from DESI), and CMB angular power spectra and type Ia supernovae (external). We briefly describe each of these datasets and their modeling.

DESI DR1 includes over 4.7 million galaxy and quasar redshifts between $0.1 < z < 2.1$, divided into several tracer classes: the Bright Galaxy Survey (BGS) from $0.1 < z< 0.4$, Luminous Red Galaxies (LRGs) in three redshift slices ($0.4 < z < 0.6$, $0.6 < z < 0.8$, $0.8 < z < 1.1$), Emission Line Galaxies (ELGs) from $1.1 < z < 1.6$, and Quasars (QSOs) from $0.8 < z < 2.1$. The power spectrum multipoles for each tracer are measured using the Feldman-Kaiser-Peacock (FKP)~\cite{Feldman_1994} estimator as implemented in \texttt{pypower}~\cite{Hand_2017}, applying weights to correct for survey selection effects and optimize two-point statistics. Small-scale systematics from fiber collisions are mitigated via a combination of angular cuts and a rotation of the data vector, window matrix, and covariance, yielding a more diagonal window function (see~\cite{DESI_II_2024}). The covariance matrix for the multipoles is estimated from $1\,000$ EZmock realizations~\cite{Forero-Sanchez_2025, Alves_inprep, Zhao_inprep} and rescaled to match the semi-empirical covariance inferred from the observed data; all known observational systematics~\cite{Findlay_2024} are incorporated. BAO measurements are extracted after reconstruction and compressed into isotropic and anisotropic components\footnote{The procedure used to extract the compressed BAO parameters is Bayesian, marginalizing over the non-BAO signal in the clustering. However, a profile likelihood study was conducted with \texttt{iminuit} and confirmed that it produced confidence intervals that differ from the Bayesian credible intervals by only a few percent for the same coverage probability, consistent with the near-Gaussian shape of the posterior distributions. Thus, we interpret the BAO measurements as giving matching frequentist and Bayesian constraints for the compressed parameters and therefore are not expected to alter our results significantly.} as described in~\cite{DESI_BAO_VI_2024}. These are calculated for the six DESI tracers and for the $\rm{Ly}\alpha$ forest~\cite{DESI_BAO_III_2024, DESI_BAO_IV_2024}. Correlations between the FS and BAO measurements are accounted for via the full mock-based covariance matrix when performing joint fits. To tighten cosmological constraints, we combine the DESI FS and BAO measurements with external probes, including the Planck ``lite" CMB likelihood~\cite{Prince_2019} and type Ia supernovae distance compilations from Union3~\cite{Union3_2023}, Dark Energy Survey (DES) Year 5~\cite{DESY5_2024} and PantheonPlus~\cite{PantheonPlus_2022}.

Table~\ref{tab:likelihoods} summarizes the datasets included in this paper. When CMB data are not included in the likelihood, we add an independent measurement of the physical baryon density, $\omega_{\rm{b}}$, from Big Bang Nucleosynthesis (BBN) and a weak constraint on the spectral index, $n_{\rm{s}}$, with $10\times$ the width of the Planck 2018 result~\cite{Planck2018_parameters}, denoted $n_{\rm{s}10}$. For the FS modeling, we work in the same ``physical" basis employed in~\cite{Zhang_2025b, DESI_FS_V_2024, DESI_FS_VII_2024} which includes galaxy bias parameters, counterterms, and stochastic parameters; we sometimes collectively refer to these as EFT nuisance parameters. The priors applied to the physical basis for the Bayesian results\footnote{These priors on the EFT nuisance parameters are only applied for the Bayesian results since the frequentist results do not apply any priors, except for BBN and $n_{\rm{s}10}$ which are treated as joint likelihoods.} are found in Table 4 from~\cite{DESI_FS_V_2024}. The conversion to the Eulerian basis, which is ultimately used in \texttt{Effort.jl} (see below) to emulate the \texttt{velocileptors} code~\cite{Maus_2025} for theory predictions, can be found in Equation~(2.2) of~\cite{Zhang_2025b}. Uniform priors are applied to cosmological parameters (except for $n_{\rm{s}}$ and $\omega_{\rm{b}}$ when CMB data are excluded) in the basis $\{ \ln{(10^{10}A_{\rm{s}})}, n_{\rm{s}}, H_0, \omega_{\rm{b}}, \omega_{\rm{c}}, w_0, w_a \}$ for the Bayesian results. The ranges for the uniform priors correspond to the boundaries chosen for our emulators\footnote{The emulator boundaries were chosen to be wide enough to encompass the relevant regions of the posteriors (to avoid results being significantly skewed from the official DESI results), but not so wide as to diminish emulator accuracy. The only exception is the full-shape only $w_0 w_a$CDM scenario, where the posterior prematurely hits the $H_0$ upper boundary, skewing the Bayesian results, which are thus excluded from our results.}, listed in Table~\ref{tab:emulator_ranges}.

\begin{table}
\centering

\renewcommand{\arraystretch}{1.1}
    \begin{tabular}{|l|c|}
    \hhline{|=|=|}
    \textbf{Parameter} & \textbf{Emulator Range} \\
    \hline
    $\ln{(10^{10}A_{\rm{s}})}$ & [2.5, 3.7]\\
    $n_{\rm{s}}$ & [0.85, 1.1] \\
    $H_0$ & [50, 85]\\
    $\omega_{\rm{b}}$ & [0.02, 0.025]\\
    $\omega_{\rm{c}}$ & [0.08, 0.16]\\
    $w_0$ & [-3, 0.5] \\
    $w_a$ & [-3, 2] \\
    \hline
    \end{tabular}

\caption{Emulator ranges for the full-shape and BAO relevant cosmological parameters.}
\label{tab:emulator_ranges}
\end{table}

\begin{table}
\centering
\resizebox{\columnwidth}{!}{
    \small 
\renewcommand{\arraystretch}{1.1}
    \begin{tabular}{|l|ll|}
    \hhline{|===|}
    \textbf{Name} & \textbf{Description} & \textbf{Ref}\\
    \hline
    DESI-FS    & DESI DR1 FS likelihood&\cite{DESI_FS_V_2024, DESI_FS_VII_2024} \\
    DESI       & Combined DESI DR1 FS$+$BAO likelihood & \cite{DESI_FS_V_2024, DESI_FS_VII_2024} \\
    \hline
    CMB        & Planck ``lite" CMB likelihood &\cite{Prince_2019}\\
    \hline
    Union3 &  Type Ia supernova likelihood from Union3 compilation&\cite{Union3_2023}\\
    DESY5 & Type Ia supernova likelihood from DES Year 5 compilation&\cite{DESY5_2024}\\
    PantheonPlus & Type Ia supernova likelihood from PantheonPlus compilation&\cite{PantheonPlus_2022}\\
    \hline
    BBN        & Independent measurement on $\omega_{\rm{b}}$ from Big Bang Nucleosynthesis, $\omega_{\rm{b}}\sim\mathcal{N}(0.02218,0.00055^2)$&\cite{Schoneberg_2024}\\
    $n_{\mathrm{s10}}$ & Weak constraint on $n_{\rm{s}}$ with width 10 times wider than \textit{Planck}, $n_{\rm{s}}\sim\mathcal{N}(0.9649,0.042^2)$&\cite{Planck2018_parameters}\\
    \hline
    \end{tabular}
}
\caption{Summary of datasets used in this analysis. The first column lists the shorthand notation for each likelihood, followed by a brief description and relevant references.}
\label{tab:likelihoods}
\end{table}

To accelerate our analysis, we employ surrogate models: specifically, \texttt{Effort.jl}~\cite{Bonici_2025} to emulate the FS power spectrum multipoles and \texttt{Capse.jl}~\cite{Bonici_2024} to model the CMB primary anisotropy power spectrum. This approach offers two main advantages: it significantly speeds up theoretical calculations, and, because these codes are differentiable~\cite{Blondel_2024}, it enables the use of gradient‐based methods. In the context of cosmological summary statistics, recent studies have demonstrated the promise of such techniques for further accelerating analysis pipelines~\cite{Campagne_2023, Piras_2023, Ruiz-Zapatero_2024, Cagliari_2025, Nygaard_2023, Balkenhol_2024, Giovanetti_2024, Balkenhol_2025, SPT-3G}. For the CMB likelihood, we adopt the compressed 2018 Planck likelihood\footnote{\href{https://github.com/JuliaCosmologicalLikelihoods/PlanckLite.jl}{\texttt{PlanckLite.jl}}}, used in~\cite{Bonici_2024} and developed in~\cite{Prince_2019}. By marginalizing over CMB‐specific nuisance parameters, this methodology reduces computational complexity. Comparisons within the CMB community show close agreement between marginalized and full likelihoods~\cite{Prince_2019, Planck_2018_CMBlikelihood, Balkenhol_2025, ACT_2025}. Finally, we ported the type Ia supernovae likelihoods\footnote{\href{https://github.com/JuliaCosmologicalLikelihoods/SNIaLikelihoods.jl}{\texttt{SNIaLikelihoods.jl}}} to \texttt{Julia} in order to use them together with the other employed codes.

\section{Methodology}
\label{sec:method}

We apply the standard profile likelihood (PL) graphical construction method~\cite{Herold_2025a,Karwal_2024,Holm_2023a,Planck_2014, Herold_2022, Campeti_2022, Holm_2023b, Nygaard_2023, Herold_2025b} to construct frequentist confidence intervals, and compare them with Bayesian credible intervals derived from the marginal posteriors. We also compare with the MAP parameters, albeit with the caveats discussed in Section~\ref{sec:intro}. We model our likelihoods and perform all optimizations and chains using the probabilistic programming language \texttt{Turing.jl}~\cite{Ge_2018, Fjelde_2025}; its compatibility with the \texttt{Julia} automatic differentiation ecosystem permits the use of gradient-based methods, speeding up convergence. We now discuss how each of the above quantities is calculated.

\subsection{Bayesian Credible Intervals}

We use the standard Markov Chain Monte Carlo (MCMC) technique to sample from the posterior distribution, defined by Bayes' theorem as: \begin{equation} \posterior \propto \likelihood \prior, \label{eq:posterior} \end{equation} where $\likelihood$ is the likelihood function and $\prior$ is the prior distribution. Specifically, we apply the ``No-U-Turn Sampling" (NUTS) Hamiltonian Monte Carlo (HMC) algorithm~\cite{Hoffman_2011}, running ten independent chains, each with $500$ burn-in steps, $2\,000$ accepted steps and a 0.65 acceptance fraction, combining to obtain a total of $20\,000$ accepted steps\footnote{To ensure sufficient convergence, we calculated the Gelman-Rubin statistic~\cite{Gelman_1992} and found that $R - 1 < 0.01$ is satisfied for all parameters.}. We obtain Bayesian credible intervals by marginalizing over the full multidimensional posterior\footnote{Given the significant computational speedup described previously, we did not use analytical marginalization over any parameters in the model.}, and then calculating the mean and corresponding uncertainty intervals using the \texttt{Python} package \texttt{GetDist}~\cite{Lewis_2019}.

\subsection{Maximum A Posteriori (MAP)}\label{sec:MAP}

To maximize the posterior distribution, as defined in Equation~\ref{eq:posterior}, we use the limited-memory Broyden-Fletcher-Goldfarb-Shanno (L-BFGS) algorithm~\cite{Liu_1989}, as implemented in \texttt{Optim.jl}~\cite{Mogensen_2018}. We run 50 independent optimizations with random starting locations throughout the parameter space to increase the chances of reaching the global maximum.

\subsection{Frequentist Confidence Intervals}

The one (two)-dimensional PLs are calculated by maximizing the likelihood over all parameters except the one (two) being profiled: \begin{equation} L(\theta_i) \equiv \max_{\theta_j, j \neq i}\likelihood \ \leftarrow \ \rm{1D \ PL} \end{equation} \begin{equation} L(\theta_i, \theta_j) \equiv \max_{\theta_k, k \neq i, j}\likelihood \ \leftarrow \ \rm{2D \ PL}. \end{equation} In the asymptotic limit of a large dataset or a Gaussian likelihood, the quantity $\deltachisquaredtheta \equiv -2 \log{(L(\theta_i)/\mathcal{L}_{\rm{max}})}$ follows a $\chi^2$ distribution with one degree of freedom (likewise $\Delta \chi^2 (\theta_i, \theta_j) \equiv -2\log{(L(\theta_i, \theta_j)/\mathcal{L}_{\rm{max}})}$ follows a $\chi^2$ distribution with two degrees of freedom). In this scenario, the 68.3 (95.4)\% one-dimensional confidence intervals correspond to $\deltachisquaredtheta < 1 \ (4)$, and the equivalent two-dimensional confidence regions correspond to $\Delta \chi^2(\theta_i, \theta_j) < 2.30 \ (6.17)$. We note that this assumption does not strictly hold given that some of our likelihoods are non-Gaussian – particularly for the $w_0w_a$ extension when constraints are weaker – and therefore our constructed confidence intervals are not guaranteed to obey exact frequentist coverage properties~\cite{Holm_2023a, Herold_2025a}. To help assess the level of disagreement, Appendix~\ref{sec:coverage} includes a coverage test for one of the dataset/parameter combinations that exhibits significant prior effects, and we find the observed coverage is statistically consistent with the expected coverage when mock power spectrum realizations are generated from an underlying $\Lambda$CDM fiducial model.

The PL is more computationally intensive than finding the MAP, as separate optimizations must be performed for each entry (parameter value) in the profile; it is important to select reasonable starting points in parameter space so that convergence can be achieved with fewer total optimizations. Since the PL peaks at the MLE by definition, we begin by calculating the MLE parameters. To do this, we apply the same procedure used to find the MAP in~\ref{sec:MAP}, but remove all Gaussian priors for the EFT nuisance parameters. Once the MLE is obtained, we use the same L-BFGS procedure to perform the optimizations for the PL, albeit in the reduced parameter space excluding the parameters being profiled. We performed between 10 and 20 independent optimizations in total for each entry\footnote{The specific number of optimizations used was based on the convergence requirements of each dataset combination, with datasets including supernovae converging with fewer optimizations and the most optimizations required when only DESI data are fit. When needed, additional optimizations were run for individual profile entries that were noisy and did not converge for the given number of minimizations.}, with the first optimization starting from the exact MLE parameters (except the parameter(s) being profiled), and the remaining runs using starting guesses centered on the MLE with random displacements added proportional to the variance of the corresponding MCMC chain. While a single optimization from the MLE point often finds the maximum, initiating multiple runs from diverse starting locations ensures a more thorough exploration of the parameter space and yields more robust estimates.

We compute 16 entries (discrete parameter values) for each one-dimensional profile and apply quadratic interpolation in log-likelihood space to produce a smooth one-dimensional profile. This ensures both that the minimum $\chi^2$ is accurate, since none of the discrete points exactly coincide with it, and allows for interpolation between entries to find the $\Delta \chi^2$ thresholds necessary to construct our confidence intervals\footnote{The parameter bounds for the profiles were selected in advance, using initial test runs, to approximately encompass the relevant $\sim \Delta \chi^2 \leq 9 \ (3 \sigma)$ regions. In most cases, the chosen resolution of 16 entries was sufficient to generate smooth interpolated profiles, without performing unnecessary computations. When necessary, additional entries were introduced to avoid unwanted features in the interpolation.}. Similar to the procedure for the one-dimensional profiles, we generate the two-dimensional profiles on a discrete $16\times16$ grid narrowly encompassing the relevant regions of the PL, and use cubic interpolation to generate the contours for the 1$\sigma$ and 2$\sigma$ confidence regions.

\section{Results}
\label{sec:results}

\begin{figure}
    \centering
    \includegraphics[width=\textwidth]{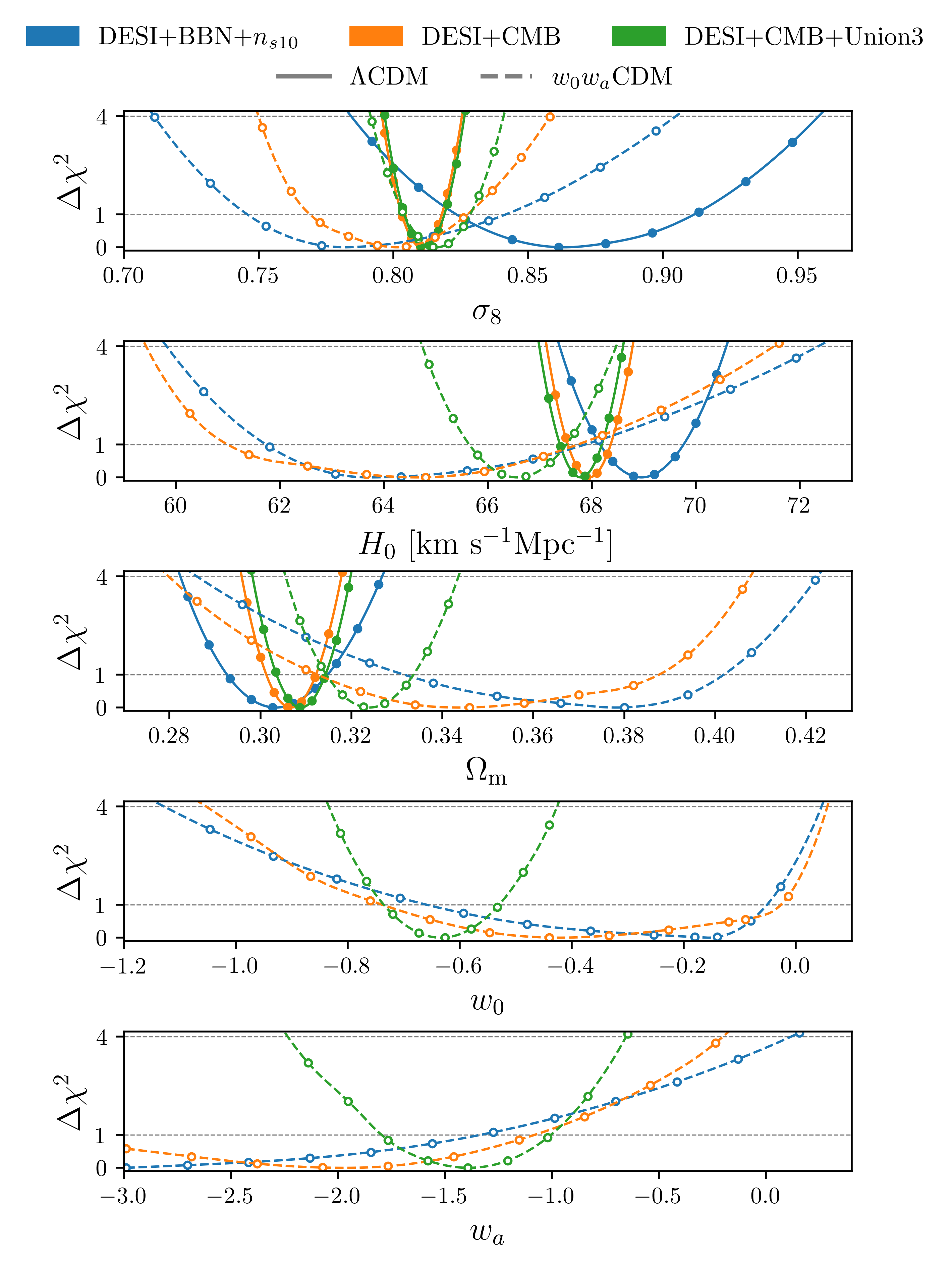}
    \caption{$\Delta \chi^2$ profiles for the profile likelihoods. The solid lines (with solid markers) indicate $\Lambda$CDM while the dashed lines (with open markers) indicate $w_0 w_a$CDM. The horizontal dashed lines indicate the $\Delta \chi^2 = 1,4$ thresholds for the 1$\sigma$ and 2$\sigma$ confidence intervals.}
    \label{fig:chi2_profiles}
\end{figure}

We compare the one-dimensional PL, marginalized posteriors (MCMC), and MAP across all dataset combinations for both the baseline $\Lambda$CDM and $w_0 w_a$CDM models. These three statistics are calculated for the following parameters: $ \{ \sigma_8, H_0, \Omega_{\rm{m}}, w_0, w_a \}$. We neglect $n_{\rm{s}}$ and $\omega_{\rm{b}}$ since they are subject to the wide $n_{\rm{s}10}$ and BBN constraints when CMB data are not included~\cite{DESI_FS_V_2024, DESI_FS_VII_2024}, as discussed previously. Given the significant interest in the question of dynamical dark energy and the apparent inconsistency with $\Lambda$CDM when DESI results are combined with CMB and supernovae datasets~\cite{DESI_BAO_VI_2024, DESI_FS_VII_2024, DESI_DR2_BAO}, we also include the two-dimensional PL for the $w_0, w_a$ parameter pair. Figure~\ref{fig:chi2_profiles} displays the $\Delta \chi^2$ results associated with the one-dimensional PLs. Since new parameters are introduced in the $w_0 w_a$CDM model, we observe wider and often asymmetric or non-Gaussian $\Delta \chi^2$ profiles compared to the $\Lambda$CDM model. We caution the reader that certain PLs for the $w_0 w_a$CDM model contain distortions, since the associated best-fit $w_a$ frequently hits the lower boundary $w_a=-3$ imposed on the model by DESI, within the range of parameter values considered in the profile. This does not affect the constructed confidence intervals if the boundary is reached when the profiled parameter is beyond the $\pm 1 \sigma$ range relative to the MLE. In some cases – specifically when supernovae are excluded – there is a non-negligible impact on the quoted confidence intervals. This effect often manifests visually as a sudden increase on one side of the $\Delta \chi^2$ profile (compared to the other side, which increases more smoothly), e.g.  the $w_0$ profiles for $\rm{DESI+BBN}+n_{\rm{s}10}$ and $\rm{DESI+CMB}$ in Figure~\ref{fig:chi2_profiles}. We note, however, that the marginalized posteriors are also impacted by the lower $w_a=-3$ boundary. To ensure maximum consistency between procedures, we still apply the boundary when computing the profiles. We now proceed to consider the $\Lambda$CDM and $w_0 w_a$CDM models individually and assess the prior effects.

\subsection{$\Lambda$CDM model}

\begin{figure}
    \centering
    \includegraphics[width=\textwidth]{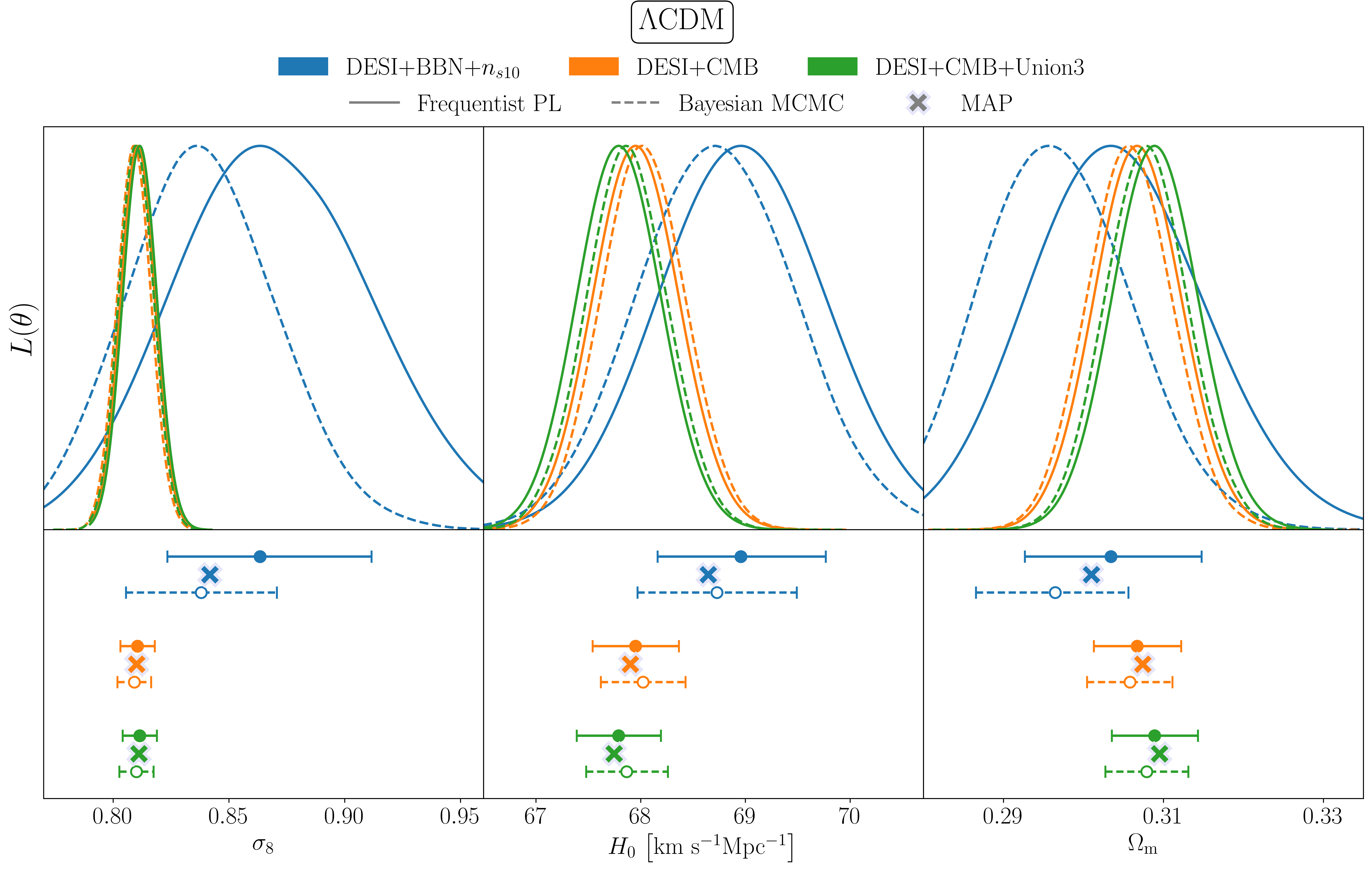}
    \caption{Top panels: profile likelihood (solid line), one-dimensional posterior (dashed line) overlaid for each of the three cosmological parameters in the $\Lambda$CDM model. Bottom panels: the corresponding 68.3\% confidence intervals from the profile likelihood (solid line), credible intervals from the one-dimensional posterior (dashed lines) and maximum a posteriori (markers).}
    \label{fig:LCDM_profiles}
\end{figure}

Figure~\ref{fig:LCDM_profiles} overlays the PL, MCMC and MAP for each dataset combination and parameter for the $\Lambda$CDM model. Table~\ref{tab:results_LCDM} summarizes these results by showing the $1 \sigma \ \rm{(68.3\%)}$ confidence intervals, along with the analogous $1\sigma$ credible intervals and MAP parameters. We observe close agreement between the PL, MCMC and MAP, with the MLE and MAP falling within the $1\sigma$ Bayesian credible intervals for all parameters and dataset combinations; the agreement becomes even stronger when additional CMB and supernovae datasets are combined with DESI data. For example, for the $\rm{DESI+BBN+}n_{\rm{s}10}$ dataset, we obtain the following frequentist 1$\sigma$ confidence intervals: 

\begin{equation}
\left.
 \begin{aligned}
\sigma_8 &= 0.863^{+0.048}_{-0.040} \\
H_0 &= (68.96^{+0.81}_{-0.80}) \,{\rm km\,s^{-1} Mpc^{-1}}\\
\Omega_{\rm{m}} &= 0.3034\pm0.0110 \\
 \end{aligned}
\ \right\}
\ \mbox{\rm{DESI+BBN+}$n_{\rm{s10}}$}\,,
\label{eqn:LCDM_frequentist_constraints1}
\end{equation} and for the $\rm{DESI+CMB+Union3}$ dataset, we obtain the following:

\begin{equation}
\left.
 \begin{aligned}
\sigma_8 &= 0.8115\pm0.0074 \\
H_0 &= (67.79\pm0.40) \,{\rm km\,s^{-1} Mpc^{-1}}\\
\Omega_{\rm{m}} &= 0.3089^{+0.0054}_{-0.0053} \\
 \end{aligned}
\ \right\}
\ \mbox{\rm{DESI+CMB+Union3}}\,.
\label{eqn:LCDM_frequentist_constraints2}
\end{equation} For $\rm{DESI+BBN+}n_{\rm{s}10}$, we observe 0.8$\sigma$, 0.3$\sigma$, 0.7$\sigma$ shifts\footnote{We use the following metric to quantify the discrepancy between the MLE and the Bayesian mean: \begin{equation} \sigma\rm{-distance} = \frac{\theta_{\rm{MLE}}-\theta_{\rm{MCMC}}}{\sigma_{\rm{MCMC}}}\,,\end{equation} where $\theta_{\rm{MLE}}$ is the MLE value, $\theta_{\rm{MCMC}}$ is the Bayesian mean and $\sigma_{\rm{MCMC}}$ is the standard deviation of the Bayesian posterior.} between the MLE and the Bayesian mean for $\sigma_8, H_0, \Omega_{\rm{m}}$, compared to only 0.2$\sigma$ shifts for the $\rm{DESI+CMB+Union3}$ dataset. The Bayesian credible intervals are narrower than the frequentist confidence intervals, which can likely be attributed, at least partially, to the weight of the EFT priors. As described in Section~\ref{sec:results}, though, these are intrinsically different quantities calculated in different ways using marginalization or profiling, so it is difficult to say with certainty where the difference arises. Overall, the close agreement between the confidence and credible intervals is indicative of minimal prior effects for the $\Lambda$CDM model, which is in agreement with the official DESI DR1 analysis~\cite{DESI_FS_VII_2024}.

\subsection{$w_0w_a$CDM model}

\begin{figure}
    \centering
    \includegraphics[width=\textwidth]{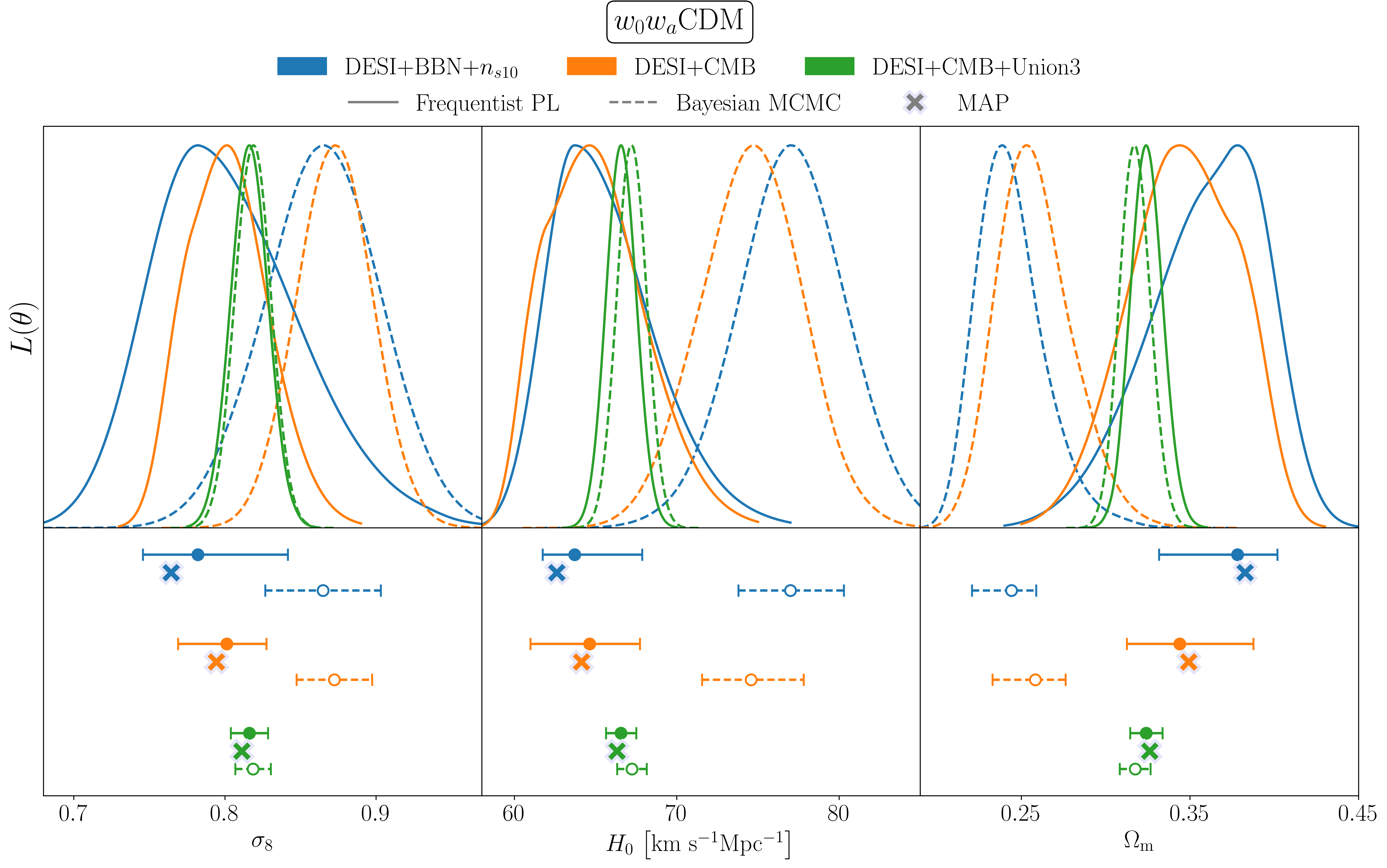}
    \includegraphics[width=0.65\textwidth]{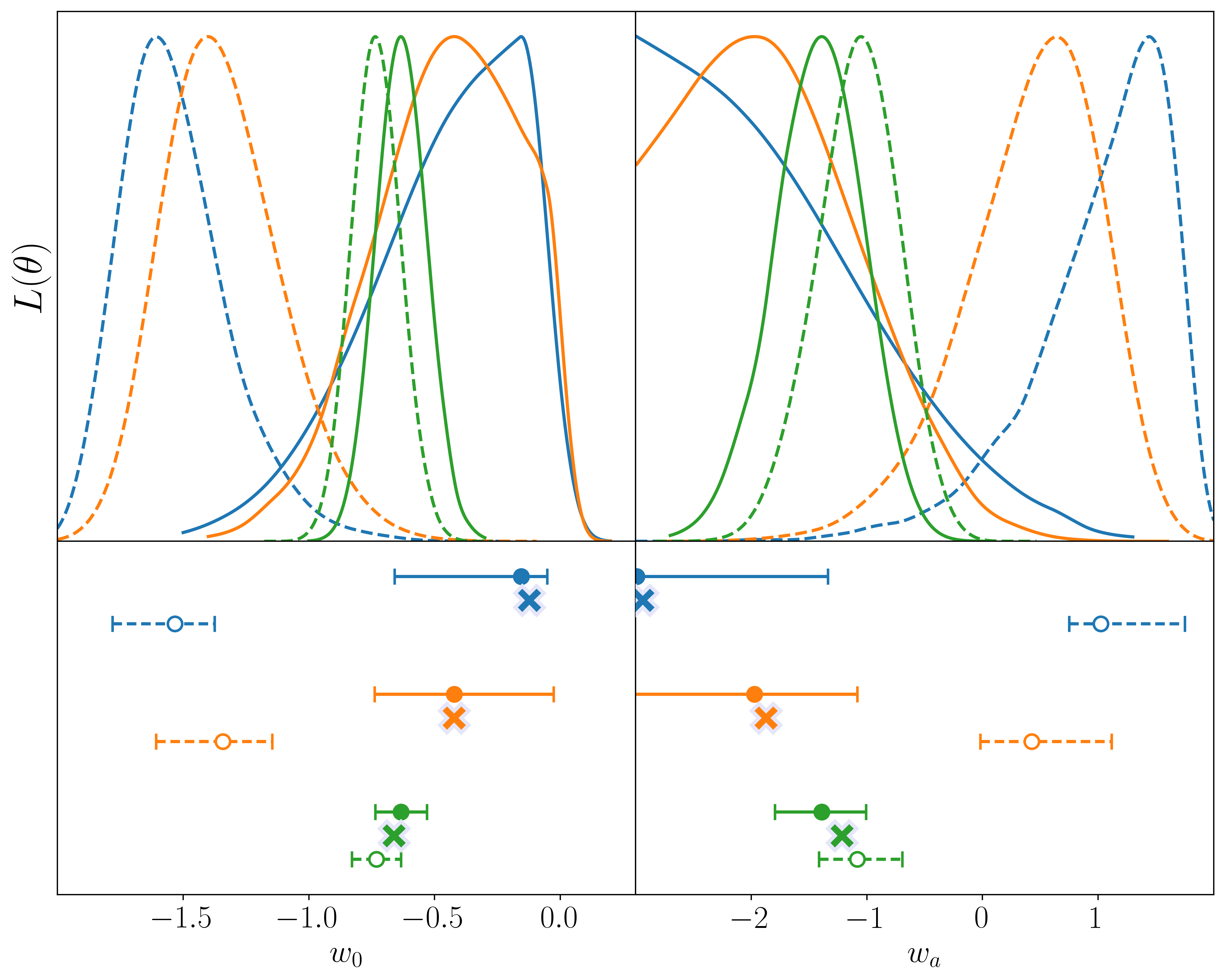}
    \caption{Top panels: profile likelihood (solid line), one-dimensional posterior (dashed line) overlaid for each of the five cosmological parameters in the $w_0 w_a$CDM model. Bottom panels: the corresponding 68.3\% confidence intervals from the profile likelihood (solid line), credible intervals from the one-dimensional posterior (dashed lines) and maximum a posteriori (markers).}
    \label{fig:w0wa_profiles}
\end{figure}

Figure~\ref{fig:w0wa_profiles} and Table~\ref{tab:results_w0waCDM} are equivalent to Figure~\ref{fig:LCDM_profiles} and Table~\ref{tab:results_LCDM}, but for the $w_0w_a$CDM extension. As expected, prior effects are significantly more pronounced for the $w_0 w_a$CDM model. While the MLE and MAP are similar, i.e. the MAP estimates fall within the 1$\sigma$ confidence intervals, for all the parameter/dataset combinations, the MLE consistently occurs well outside the 1$\sigma$ Bayesian credible intervals. The notable exceptions occur when supernovae are included, as they help break degeneracies among the background expansion parameters, tightening constraints sufficiently to reduce prior effects. For the $\rm{DESI+BBN+}n_{\rm{s}10}$ dataset, for example, we obtain the following frequentist $1\sigma$ confidence intervals:

\begin{equation}
\left.
 \begin{aligned}
\sigma_8 &= 0.782^{+0.060}_{-0.036} \\
H_0 &= (63.7^{+4.2}_{-2.0}) \,{\rm km\,s^{-1} Mpc^{-1}}\\
\Omega_{\rm{m}} &= 0.378^{+0.024}_{-0.047} \\
w_0 &= -0.16^{+0.10}_{-0.50}\\
w_a &= -3.0^{+1.7}_{}\\
 \end{aligned}
\ \right\}
\ \mbox{\rm{DESI+BBN+}$n_{\rm{s10}}$}\,,
\label{eqn:w0waCDM_frequentist_constraints1}
\end{equation} and for the $\rm{DESI+CMB+Union3}$ dataset, we obtain the following:

\begin{equation}
\left.
 \begin{aligned}
\sigma_8 &= 0.816\pm0.012 \\
H_0 &= (66.57^{+0.94}_{-0.92}) \,{\rm km\,s^{-1} Mpc^{-1}}\\
\Omega_{\rm{m}} &= 0.3240^{+0.0097}_{-0.0095} \\
w_0 &= -0.634 \pm 0.100\\
w_a &= -1.39^{+0.38}_{-0.41}\\
 \end{aligned}
\ \right\}
\ \mbox{\rm{DESI+CMB+Union3}}\,.
\label{eqn:w0waCDM_frequentist_constraints2}
\end{equation} For $\rm{DESI+BBN+}n_{\rm{s}10}$, we observe 2.1$\sigma$, 4.1$\sigma$, 6.5$\sigma$, 6.3$\sigma$, 6.6$\sigma$ shifts between the MLE and the Bayesian mean for $\sigma_8$, $H_0$, $\Omega_{\rm{m}}$, $w_0$, $w_a$, compared to only 0.2$\sigma$, 0.7$\sigma$, 0.7$\sigma$, 1.0$\sigma$, 0.9$\sigma$ shifts for the $\rm{DESI+CMB+Union3}$ dataset. While a reduction was also observed for $\Lambda$CDM, none of the deviations were more than $1\sigma$, while for $w_0w_a$CDM, deviations go from being statistically significant to less than $1 \sigma$ when CMB and supernovae data are added.

\begin{figure}
    \centering
    \includegraphics[width=0.93\textwidth]{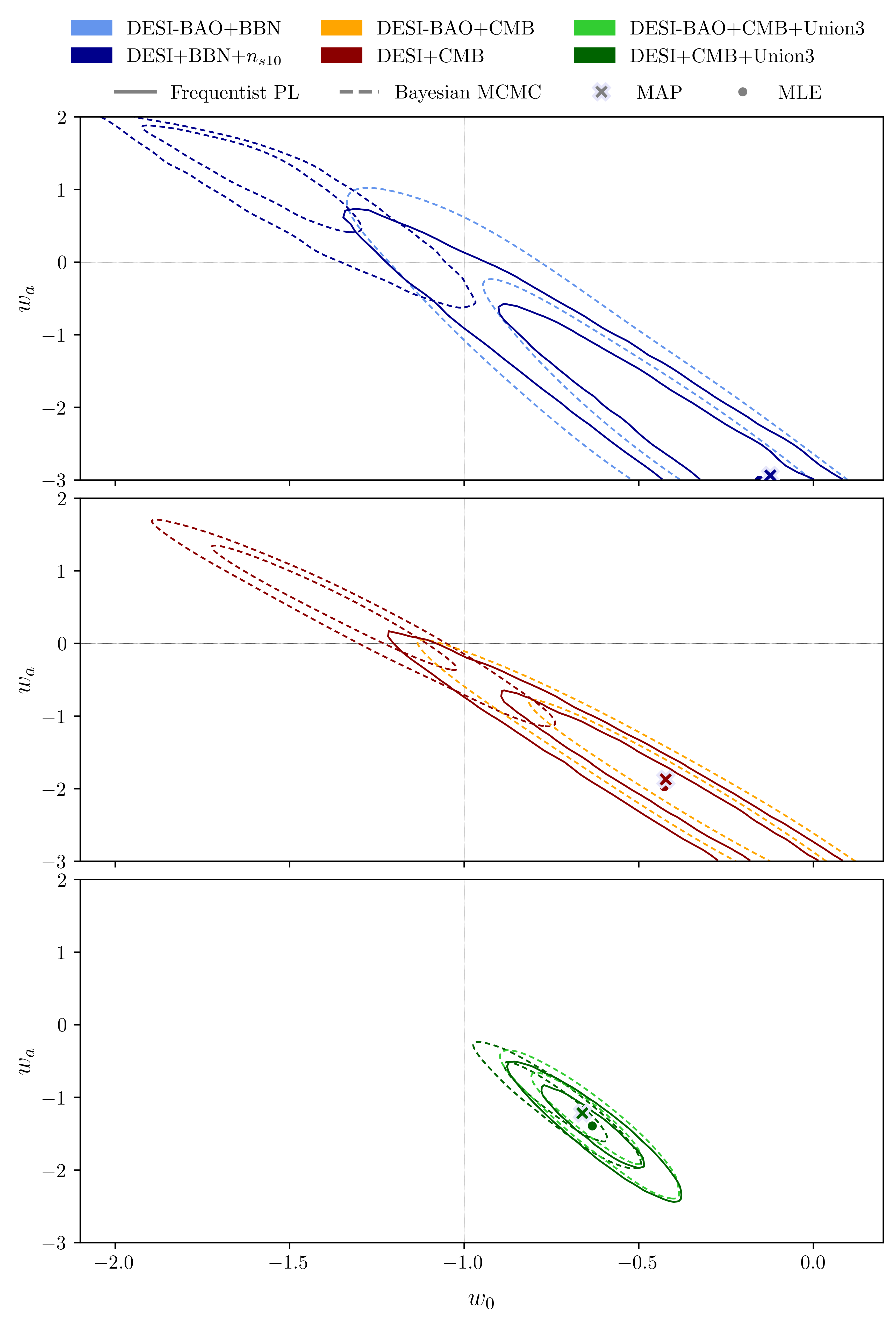}
    \caption{Two-dimensional 1$\sigma$ and 2$\sigma$ frequentist confidence regions (solid line) from the two-dimensional profile likelihood and the maximum a posteriori from the posterior (marker) overlaid on the corresponding Bayesian credible regions (dashed line) for the $w_0w_a$ parameter pair for the various dataset combinations. The BAO-only Bayesian credible regions (lighter dashed line) are overlaid on each dataset for comparison.}
    \label{fig:2D_profiles}
\end{figure}

Lastly, Figure~\ref{fig:2D_profiles} overlays the 1$\sigma$ and 2$\sigma$ two-dimensional confidence regions (from the two-dimensional PL), two-dimensional credible regions (from the two-dimensional marginal posterior) and the MAP parameters. We reach similar qualitative conclusions as found for the one-dimensional scenarios; the MAP occurs within the 1$\sigma$ confidence regions, while the MLE falls outside the 1$\sigma$ credible regions unless supernovae are included. Importantly though, we observe the same degeneracy directions in both the frequentist and Bayesian approaches.

While the existing DESI Bayesian constraints are largely insensitive to prior effects for $\Lambda$CDM, the $w_0w_a$CDM constraints are not. Having now performed a frequentist analysis where the prior effects are suppressed, we can make several comparisons that were not possible before. As presented in Table~\ref{tab:results_w0waCDM}, the frequentist constraints are largely consistent between full-shape only and joint full-shape/BAO fits, indicating similar parameter preferences for both full-shape and BAO probes. This is in sharp contrast to the Bayesian framework, where statistically significant discrepancies exist depending on whether or not full-shape data are included. To highlight this, in Figure~\ref{fig:2D_profiles} we also overlay the two-dimensional marginal posteriors for the analogous BAO-only models to highlight the minimal prior effects present when full-shape data are excluded\footnote{The BAO-only chains are taken from the official DESI results (as opposed to running chains under our pipeline). While they serve as a useful comparison, we note that they use a different CMB likelihood and slightly different priors/parameterizations. However, given the reduced parameter space for BAO-only modeling, this is not expected to significantly alter the results.}. Unlike the Bayesian results, where the $w_0w_a$ contours only move to the quadrant favored by the BAO results once supernovae are added, the frequentist constraints for all data combinations are found in the same quadrant and are largely consistent with the BAO-only Bayesian constraints. These key observations collectively highlight the importance of accounting for prior effects when obtaining parameter constraints, particularly for the $w_0w_a$CDM extension.

\begin{table}
\centering
\resizebox{\columnwidth}{!}{
    \small 
\setcellgapes{3pt}\makegapedcells  
\renewcommand{\arraystretch}{1.9}
    \begin{tabular}{|l|ccc|}
    \hhline{|====|}
    \bf{Model/Dataset} & $\sigma_8$ & $H_0 \ \rm{[km\,s^{-1}Mpc^{-1}]}$ & $\Omega_{\mathrm{m}}$ \\[0.1cm]
    \hline
    {\bf Flat} $\boldsymbol{\Lambda}${\bf CDM} &&&\\
    
    \makecell[l]{DESI-FS+BBN+$n_{\mathrm{s10}}$ (PL)} & $0.861^{+0.042}_{-0.041}$ & $70.4\pm1.1$ & $0.300^{+0.014}_{-0.013}$ \\
    \makecell[l]{DESI-FS+BBN+$n_{\mathrm{s10}}$ (MCMC)}  & $0.837\pm0.034$ & $70.2\pm1.0$ & $0.284^{+0.011}_{-0.012}$ \\
    \makecell[l]{DESI-FS+BBN+$n_{\mathrm{s10}}$ (MAP)} & $0.839$ & $69.9$ & $0.291$ \\
   
    \hdashline
    
    \makecell[l]{DESI+BBN+$n_{\mathrm{s10}}$ (PL)} & $0.863^{+0.048}_{-0.040}$ & $68.96^{+0.81}_{-0.80}$ & $0.3034\pm0.0110$  \\
    \makecell[l]{DESI+BBN+$n_{\mathrm{s10}}$ (MCMC)} & $0.838^{+0.033}_{-0.032}$ & $68.73\pm0.76$ & $0.2965^{+0.0091}_{-0.0099}$  \\
    \makecell[l]{DESI+BBN+$n_{\mathrm{s10}}$ (MAP)} & $0.842$ & $68.64$ & $0.3010$  \\

    \hdashline
    
    DESI+CMB (PL) & $0.8105^{+0.0075}_{-0.0074}$ & $67.95\pm0.41$ & $0.3067^{+0.0055}_{-0.0054}$ \\
    DESI+CMB (MCMC) & $0.8091\pm0.0073$ & $68.02\pm0.40$ & $0.3058^{+0.0053}_{-0.0054}$  \\
    DESI+CMB (MAP) & $0.8101$ & $67.90$ & $0.3074$  \\

    \hdashline
    
    DESI+CMB+U3 (PL) & $0.8115\pm0.0074$ & $67.79\pm0.40$ & $0.3089^{+0.0054}_{-0.0053}$  \\
    DESI+CMB+U3 (MCMC) & $0.8100\pm0.0074$ & $67.87\pm0.39$ & $0.3079\pm0.0052$  \\
    DESI+CMB+U3 (MAP) & $0.8111$ & $67.74$ & $0.3095$  \\

    \hdashline
    
    DESI+CMB+D5 (PL) & $0.8126\pm0.0074$ & $67.60\pm0.39$ & $0.3115^{+0.0053}_{-0.0052}$  \\
    DESI+CMB+D5 (MCMC) & $0.8112\pm0.0073$ & $67.68^{+0.39}_{-0.38}$ & $0.3105\pm0.0052$  \\
    DESI+CMB+D5 (MAP) & $0.8122$ & $67.56$ & $0.3120$  \\

    \hdashline 
    
    DESI+CMB+PP (PL) & $0.8114\pm0.0074$ & $67.79\pm0.39$ & $0.3089^{+0.0053}_{-0.0052}$  \\
    DESI+CMB+PP (MCMC) & $0.8101\pm0.0073$ & $67.86^{+0.39}_{-0.38}$ & $0.3080\pm0.0051$  \\
    DESI+CMB+PP (MAP) & $0.8111$ & $67.75$ & $0.3095$  \\
    \hline
    \end{tabular}
}
\caption{68.3\% confidence intervals for frequentist profile likelihoods, 68.3\% credible intervals for Bayesian MCMC chains, and MAP parameters for each dataset/parameter combination for the $\Lambda$CDM model. To save space, Union3, DESY5, PantheonPlus are denoted as U3, D5, PP.
}

\label{tab:results_LCDM}
\end{table}
\begin{table}
\centering
\resizebox{\columnwidth}{!}{
    \small 
\setcellgapes{3pt}\makegapedcells  
\renewcommand{\arraystretch}{2.7}
    \begin{tabular}{|l|ccccc|}
    \hhline{|======|}
    \bf{Model/Dataset} & $\sigma_8$ & $H_0 \ \rm{[km\,s^{-1}Mpc^{-1}]}$ & $\Omega_{\mathrm{m}}$ & $w_0$ & $w_a$ \\[0.1cm]
    \hline
    {\bf Flat} $\boldsymbol{w_0 w_a}${\bf CDM} &&&&&\\
    
    \makecell[l]{DESI-FS+BBN+$n_{\mathrm{s10}}$ (PL)} & $0.801^{+0.067}_{-0.055}$& $66.1^{+5.9}_{-4.0}$& $0.346^{+0.049}_{-0.056}$& $-0.42^{+0.39}_{-0.62}$& $-2.1^{+2.0}$\\
    \makecell[l]{DESI-FS+BBN+$n_{\mathrm{s10}}$ (MAP)} & $0.786$& $65.7$& $0.335$& $-0.43$& $-2.0$\\

    \hdashline
    
    \makecell[l]{DESI+BBN+$n_{\mathrm{s10}}$ (PL)} & $0.782^{+0.060}_{-0.036}$ & $63.7^{+4.2}_{-2.0}$& $0.378^{+0.024}_{-0.047}$ & $-0.16^{+0.10}_{-0.50}$ & $-3.0^{+1.7}$\\
    \makecell[l]{DESI+BBN+$n_{\mathrm{s10}}$ (MCMC)} & $0.865\pm0.038$ & $77.0^{+3.3}_{-3.2}$& $0.244^{+0.015}_{-0.023}$ & $-1.53^{+0.16}_{-0.25}$ & $1.02^{+0.73}_{-0.27}$\\
    \makecell[l]{DESI+BBN+$n_{\mathrm{s10}}$ (MAP)} & $0.764$ & $62.6$& $0.383$ & $-0.12$ & $-2.94$\\

    \hdashline
    
    DESI+CMB (PL) & $0.801^{+0.026}_{-0.032}$ & $64.6^{+3.1}_{-3.6}$& $0.344^{+0.044}_{-0.031}$ & $-0.42^{+0.40}_{-0.32}$ & $-1.97^{+0.89}$\\
    DESI+CMB (MCMC) & $0.872\pm0.025$ & $74.6^{+3.2}_{-3.0}$& $0.258^{+0.018}_{-0.025}$ & $-1.34^{+0.20}_{-0.27}$ & $0.43^{+0.69}_{-0.45}$\\
    DESI+CMB (MAP) & $0.794$ & $64.1$& $0.349$ & $-0.42$ & $-1.87$\\

    \hdashline
    
    DESI+CMB+U3 (PL) & $0.816\pm0.012$ & $66.57^{+0.94}_{-0.92}$ & $0.3240^{+0.0097}_{-0.0095}$ & $-0.634\pm0.100$ & $-1.39^{+0.38}_{-0.41}$ \\
    DESI+CMB+U3 (MCMC) & $0.819\pm0.012$ & $67.23\pm0.91$ & $0.3174\pm0.0091$ & $-0.731\pm0.098$ & $-1.08^{+0.39}_{-0.33}$ \\
    DESI+CMB+U3 (MAP) & $0.811$ & $66.31$ & $0.3260$ & $-0.662$ & $-1.22$ \\

    \hdashline
    
    DESI+CMB+D5 (PL) & $0.821\pm0.011$ & $67.26^{+0.65}_{-0.64}$ & $0.3171^{+0.0067}_{-0.0065}$ & $-0.715^{+0.071}_{-0.068}$ & $-1.15^{+0.31}_{-0.33}$ \\
    DESI+CMB+D5 (MCMC) & $0.820\pm0.011$ & $67.40\pm0.62$ & $0.3157^{+0.0064}_{-0.0063}$ & $-0.756\pm0.065$ & $-1.01^{+0.31}_{-0.27}$ \\
    DESI+CMB+D5 (MAP) & $0.816$ & $67.02$ & $0.3190$ & $-0.741$ & $-0.99$ \\

    \hdashline
    
    DESI+CMB+PP (PL) & $0.826\pm0.012$ & $68.10\pm0.70$ & $0.3089^{+0.0068}_{-0.0067}$ & $-0.820^{+0.065}_{-0.063}$ & $-0.83^{+0.27}_{-0.30}$ \\
    DESI+CMB+PP (MCMC) & $0.825\pm0.011$ & $68.27\pm0.68$ & $0.3073^{+0.0065}_{-0.0064}$ & $-0.855\pm0.061$ & $-0.71^{+0.28}_{-0.25}$ \\
    DESI+CMB+PP (MAP) & $0.821$ & $67.82$ & $0.3111$ & $-0.839$ & $-0.70$ \\
    \hline
    \end{tabular}
}
\caption{68.3\% confidence intervals for frequentist profile likelihoods, 68.3\% credible intervals for Bayesian MCMC chains, and MAP parameters for each dataset/parameter combination for the $w_0 w_a$CDM model. The MCMC results for full-shape only are excluded as the projection effects for $H_0$ are severe and the emulator boundary is hit without encompassing the full posterior.
}

\label{tab:results_w0waCDM}
\end{table}

\section{Conclusions}
\label{sec:summary}

We have presented a frequentist inference of the DESI DR1 full-shape measurements~\cite{DESI_FS_V_2024, DESI_FS_VII_2024} and derived cosmological constraints under both $\Lambda$CDM and $w_0w_a$CDM models. For $\Lambda$CDM, our analysis yields confidence intervals of $\sigma_8 = 0.863^{+0.048}_{-0.040}$, $H_0 = 68.96^{+0.81}_{-0.80} \ \rm{km \ s^{-1}Mpc^{-1}}$, and $\Omega_{\rm m} = 0.3034\pm0.0110$, when using DESI+BBN+$n_\mathrm{s10}$. In the $w_0w_a$CDM case, the prior independence of the frequentist method mitigates the prior effects seen in Bayesian analyses. We obtain: $\sigma_8 = 0.782^{+0.060}_{-0.036}$, $H_0 = 63.7^{+4.2}_{-2.0} \ \rm{km \ s^{-1}Mpc^{-1}}$, $\Omega_{\rm m} = 0.378^{+0.024}_{-0.047}$, $w_0 = -0.16^{+0.10}_{-0.50}$ and $w_a = -3.0^{+1.7}$ from DESI+BBN+$n_\mathrm{s10}$. The constraints tighten when CMB data are included, but the prior effects are only significantly reduced with the addition of supernovae data.

Unlike Bayesian inferences, which are susceptible to prior effects when working in poorly-constrained nuisance parameter spaces, frequentist inferences make no assumptions about priors. The objective of this study was to compare these two approaches in order to study the impact of prior choices. As in the baseline DESI analysis, we observed limited prior effects for the $\Lambda$CDM model, with the MLE and MAP falling within the 1$\sigma$ credible intervals. By comparison, despite the MAP falling within the 1$\sigma$ confidence intervals for the $w_0w_a$CDM model, prior effects were very pronounced with the MLE almost always falling beyond the $1\sigma$ credible intervals; only when supernovae were combined with DESI and CMB data did the prior effects reduce significantly due to their degeneracy-breaking ability. These findings, together with previous PL analyses, demonstrate the importance of carefully considering prior effects when performing full-shape analyses for extended cosmological models such as $w_0w_a$CDM. Complementing the standard Bayesian approach with the prior-independent frequentist interpretation of uncertainty provides a more nuanced interpretation of results, less susceptible to misleading conclusions.

We also emphasize the comparatively limited computational resources required to generate our results, given both the speedup provided by the use of emulators that avoid repeated manual computations of cosmological observables~\cite{Bonici_2024, Bonici_2025} and the use of gradient-based minimizers, which improve convergence~\cite{Liu_1989, Nygaard_2023}. Most dataset/model combinations only require a couple of minutes to reach convergence for individual minimizations on the Digital Alliance Narval supercomputer. Given the 10-20 optimizations performed for each individual entry and 16 total entries per profile, our PL computations (across all dataset/model/parameter combinations) collectively required $\sim$ 1000 CPU hours. The MAP/MLE calculations required fewer resources (collectively $\sim$ 100 CPU hours), since there are not multiple profile entries, each requiring their own computations. Consequently, the results in this paper can be reproduced at a reasonably low computational cost. 

We are currently working on a follow-up analysis for DESI DR2 to study how strongly prior effects are suppressed with the increased quantity of data. This includes significant improvements to the speed of our code, which will allow us to explore alternative approaches such as the profiled Feldman-Cousins method~\cite{Acero_2025}, which are more computationally intensive but provide alternative mechanisms for defining confidence regions. We will also provide tests to verify the coverage properties of both approaches. Preliminary results suggest that the coverage probabilities associated with the confidence intervals resulting from the profile likelihood method are close to those in the current work. This will be discussed in further detail in our follow-up analysis.

\section{Data Availability}
The data used in this analysis are public along with Data Release 1 (details in \url{https://data.desi.lbl.gov/doc/releases/}). The data points corresponding to the figures in this paper are available at \url{https://doi.org/10.5281/zenodo.16876979}.

\acknowledgments
JM acknowledges support from the Ontario Graduate Scholarship program through the Ontario Ministry of Colleges and Universities, and from the Canada Graduate Research Scholarship Doctoral program through the Natural Sciences and Engineering Research Council of Canada (NSERC).
WP acknowledges the support of the NSERC, [funding reference number RGPIN-2025-03931] and from the Canadian Space Agency.
Research at Perimeter Institute is supported in part by the Government of Canada through the Department of Innovation, Science and Economic Development Canada and by the Province of Ontario through the Ministry of Colleges and Universities.
This research was enabled in part by support provided by Compute Ontario (computeontario.ca) and the Digital Research Alliance of Canada (alliancecan.ca).

This material is based upon work supported by the U.S. Department of Energy (DOE), Office of Science, Office of High-Energy Physics, under Contract No. DE–AC02–05CH11231, and by the National Energy Research Scientific Computing Center, a DOE Office of Science User Facility under the same contract. Additional support for DESI was provided by the U.S. National Science Foundation (NSF), Division of Astronomical Sciences under Contract No. AST-0950945 to the NSF’s National Optical-Infrared Astronomy Research Laboratory; the Science and Technology Facilities Council of the United Kingdom; the Gordon and Betty Moore Foundation; the Heising-Simons Foundation; the French Alternative Energies and Atomic Energy Commission (CEA); the Secretariat of Science, Humanities, Technology and Innovation (SECIHTI) of Mexico; the Ministry of Science, Innovation and Universities of Spain (MICIU/AEI/10.13039/501100011033), and by the DESI Member Institutions: \url{https://www.desi.lbl.gov/collaborating-institutions}. Any opinions, findings, and conclusions or recommendations expressed in this material are those of the author(s) and do not necessarily reflect the views of the U. S. National Science Foundation, the U. S. Department of Energy, or any of the listed funding agencies.

The authors are honored to be permitted to conduct scientific research on I'oligam Du'ag (Kitt Peak), a mountain with particular significance to the Tohono O’odham Nation.

\appendix

\section{Coverage Test}
\label{sec:coverage}
As noted previously, the profile likelihood (PL) method only guarantees correct frequentist coverage under asymptotic conditions, i.e. a Gaussian likelihood, which are not satisfied in several scenarios considered here. Nevertheless, in cases where there are strong prior effects, the PL-based confidence intervals may exhibit better frequentist coverage than Bayesian credible intervals if the latter are interpreted in a frequentist sense. We perform an explicit coverage test for a case known to have severe prior effects: the joint full-shape/BAO constraint of $H_0$. We assume a fiducial ``true" model\footnote{We adopt the Planck 2018 $\Lambda$CDM model for cosmological parameters and fix the EFT parameters to the best-fit values obtained for our most constraining dataset (joint full-shape/BAO/CMB/supernovae under $\Lambda$CDM).} and generate 100 mock realizations by adding multivariate Gaussian noise, using the data covariance matrices, to the noiseless theory prediction computed from our emulators. For each realization, we compute 1$\sigma$ confidence intervals (from the profile likelihood) and 1$\sigma$ credible intervals (from the marginalized posterior), for both the $\Lambda$CDM and $w_0w_a$CDM models. Coverage is then evaluated as the fraction of intervals containing the true value.

\begin{figure}
    \centering
    \includegraphics[width=0.9\textwidth]{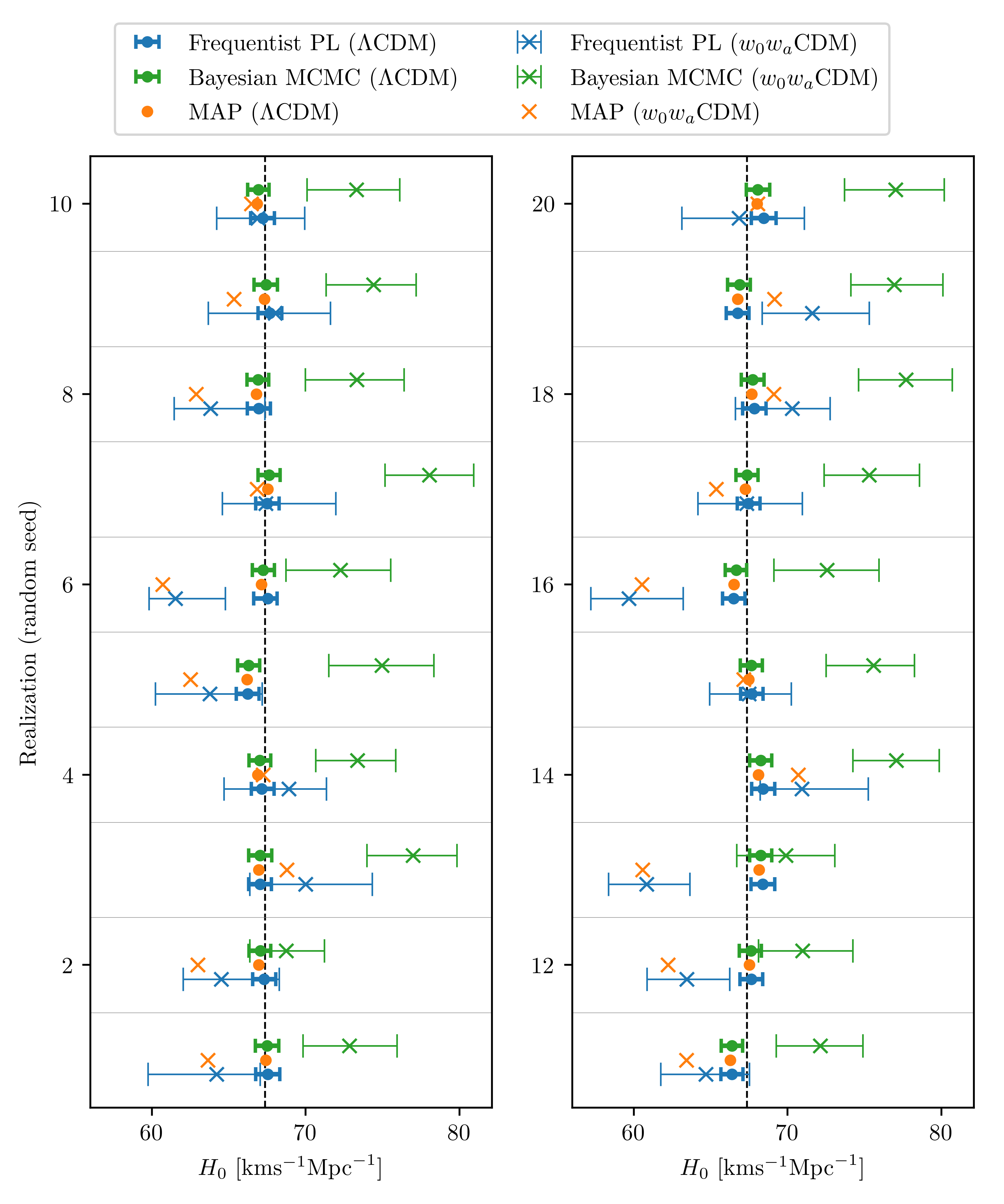}
    \caption{The confidence intervals (blue), credible intervals (green) and MAP (orange) when constraining $H_0$ in the $\Lambda$CDM (circle marker) and $w_0w_a$CDM (cross marker) models, for the first 20 out of 100 mock realizations (the vertical axis denotes the random seeds) of the joint full-shape power spectrum/BAO parameter vector. The vertical dashed line shows the true underlying value of $H_0$ (the Planck 2018 value).}
    \label{fig:freq_bay_coverage}
\end{figure}

\begin{figure}
    \centering
    \includegraphics[width=0.9\textwidth]{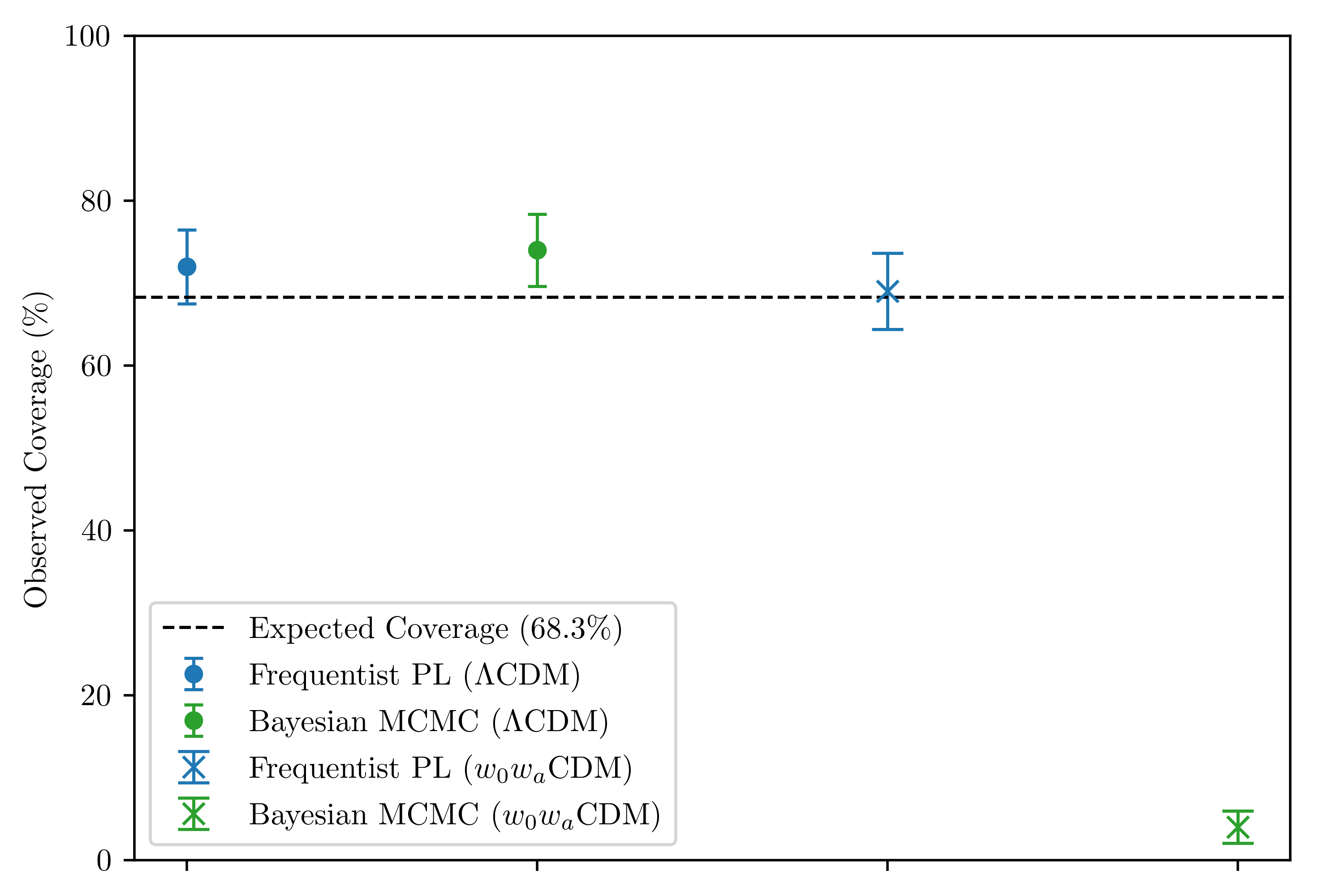}
    \caption{Observed coverage percentages and associated uncertainties when interpreting various intervals as frequentist confidence intervals for $\Lambda$CDM (circles) and $w_0w_a$CDM (crosses). Blue symbols show the intervals coming from the PL approach, while green symbols show the coverage if interpreting Bayesian credible intervals as frequentist confidence intervals.}
    \label{fig:coverage_percent}
\end{figure}

Figure~\ref{fig:freq_bay_coverage} shows the results for the first 20 out of 100 independent realizations\footnote{We opt not to show all 100 realizations as the plot becomes too crowded to clearly visualize each realization.} (generated using distinct random seeds). Consistent with the results throughout this work, we find close agreement between the frequentist confidence intervals and Bayesian credible intervals for $\Lambda$CDM. In contrast, for the $w_0w_a$CDM model the credible intervals for $H_0$ are systematically biased high and rarely contain the truth despite the MAP remaining well within the PL confidence intervals in both models. Figure~\ref{fig:coverage_percent} summarizes the results in terms of observed coverage and associated binomial uncertainties, $\sqrt{p(1-p)/N}$, where $p$ is the observed coverage fraction and $N=100$. We find that the coverage of PL confidence intervals is consistent with the expected 68.3\% level for both $\Lambda$CDM and $w_0w_a$CDM, with similar behavior for the credible intervals in $\Lambda$CDM (with tighter agreement for the former). However, the credible intervals for $w_0w_a$CDM when interpreted as confidence intervals show dramatic undercoverage. We attribute this primarily to the effect of the prior, which is substantially stronger in this model. Residual deviations in the other cases are likely due to a combination of intrinsic error (finite number of samples), emulator noise, and expected violations of asymptotic assumptions. Overall, these results indicate that the PL-based confidence intervals do not exhibit significant over- or under-coverage, even in the presence of strong prior effects.

\section{Author Affiliations}
\label{sec:affiliations}
\noindent \hangindent=.5cm $^{1}${Department of Physics and Astronomy, University of Waterloo, 200 University Ave W, Waterloo, ON N2L 3G1, Canada}

\noindent \hangindent=.5cm $^{2}${Waterloo Centre for Astrophysics, University of Waterloo, 200 University Ave W, Waterloo, ON N2L 3G1, Canada}

\noindent \hangindent=.5cm $^{3}${Perimeter Institute for Theoretical Physics, 31 Caroline St. North, Waterloo, ON N2L 2Y5, Canada}

\noindent \hangindent=.5cm $^{4}${Lawrence Berkeley National Laboratory, 1 Cyclotron Road, Berkeley, CA 94720, USA}

\noindent \hangindent=.5cm $^{5}${Department of Physics, Boston University, 590 Commonwealth Avenue, Boston, MA 02215 USA}

\noindent \hangindent=.5cm $^{6}${Dipartimento di Fisica ``Aldo Pontremoli'', Universit\`a degli Studi di Milano, Via Celoria 16, I-20133 Milano, Italy}

\noindent \hangindent=.5cm $^{7}${INAF-Osservatorio Astronomico di Brera, Via Brera 28, 20122 Milano, Italy}

\noindent \hangindent=.5cm $^{8}${Department of Physics \& Astronomy, University College London, Gower Street, London, WC1E 6BT, UK}

\noindent \hangindent=.5cm $^{9}${Institut d'Estudis Espacials de Catalunya (IEEC), c/ Esteve Terradas 1, Edifici RDIT, Campus PMT-UPC, 08860 Castelldefels, Spain}

\noindent \hangindent=.5cm $^{10}${Institute of Space Sciences, ICE-CSIC, Campus UAB, Carrer de Can Magrans s/n, 08913 Bellaterra, Barcelona, Spain}

\noindent \hangindent=.5cm $^{11}${Institute for Computational Cosmology, Department of Physics, Durham University, South Road, Durham DH1 3LE, UK}

\noindent \hangindent=.5cm $^{12}${Instituto de F\'{\i}sica, Universidad Nacional Aut\'{o}noma de M\'{e}xico,  Circuito de la Investigaci\'{o}n Cient\'{\i}fica, Ciudad Universitaria, Cd. de M\'{e}xico  C.~P.~04510,  M\'{e}xico}

\noindent \hangindent=.5cm $^{13}${IRFU, CEA, Universit\'{e} Paris-Saclay, F-91191 Gif-sur-Yvette, France}

\noindent \hangindent=.5cm $^{14}${Department of Astronomy \& Astrophysics, University of Toronto, Toronto, ON M5S 3H4, Canada}

\noindent \hangindent=.5cm $^{15}${Department of Physics \& Astronomy and Pittsburgh Particle Physics, Astrophysics, and Cosmology Center (PITT PACC), University of Pittsburgh, 3941 O'Hara Street, Pittsburgh, PA 15260, USA}

\noindent \hangindent=.5cm $^{16}${University of California, Berkeley, 110 Sproul Hall \#5800 Berkeley, CA 94720, USA}

\noindent \hangindent=.5cm $^{17}${Institut de F\'{i}sica d’Altes Energies (IFAE), The Barcelona Institute of Science and Technology, Edifici Cn, Campus UAB, 08193, Bellaterra (Barcelona), Spain}

\noindent \hangindent=.5cm $^{18}${Departamento de F\'isica, Universidad de los Andes, Cra. 1 No. 18A-10, Edificio Ip, CP 111711, Bogot\'a, Colombia}

\noindent \hangindent=.5cm $^{19}${Observatorio Astron\'omico, Universidad de los Andes, Cra. 1 No. 18A-10, Edificio H, CP 111711 Bogot\'a, Colombia}

\noindent \hangindent=.5cm $^{20}${Institute of Cosmology and Gravitation, University of Portsmouth, Dennis Sciama Building, Portsmouth, PO1 3FX, UK}

\noindent \hangindent=.5cm $^{21}${University of Virginia, Department of Astronomy, Charlottesville, VA 22904, USA}

\noindent \hangindent=.5cm $^{22}${Fermi National Accelerator Laboratory, PO Box 500, Batavia, IL 60510, USA}

\noindent \hangindent=.5cm $^{23}${Steward Observatory, University of Arizona, 933 N. Cherry Avenue, Tucson, AZ 85721, USA}

\noindent \hangindent=.5cm $^{24}${Center for Cosmology and AstroParticle Physics, The Ohio State University, 191 West Woodruff Avenue, Columbus, OH 43210, USA}

\noindent \hangindent=.5cm $^{25}${Department of Physics, The Ohio State University, 191 West Woodruff Avenue, Columbus, OH 43210, USA}

\noindent \hangindent=.5cm $^{26}${The Ohio State University, Columbus, 43210 OH, USA}

\noindent \hangindent=.5cm $^{27}${Department of Physics, University of Michigan, 450 Church Street, Ann Arbor, MI 48109, USA}

\noindent \hangindent=.5cm $^{28}${University of Michigan, 500 S. State Street, Ann Arbor, MI 48109, USA}

\noindent \hangindent=.5cm $^{29}${Department of Physics, The University of Texas at Dallas, 800 W. Campbell Rd., Richardson, TX 75080, USA}

\noindent \hangindent=.5cm $^{30}${NSF NOIRLab, 950 N. Cherry Ave., Tucson, AZ 85719, USA}

\noindent \hangindent=.5cm $^{31}${Department of Physics, Southern Methodist University, 3215 Daniel Avenue, Dallas, TX 75275, USA}

\noindent \hangindent=.5cm $^{32}${Department of Physics and Astronomy, University of California, Irvine, 92697, USA}

\noindent \hangindent=.5cm $^{33}${Sorbonne Universit\'{e}, CNRS/IN2P3, Laboratoire de Physique Nucl\'{e}aire et de Hautes Energies (LPNHE), FR-75005 Paris, France}

\noindent \hangindent=.5cm $^{34}${Departament de F\'{i}sica, Serra H\'{u}nter, Universitat Aut\`{o}noma de Barcelona, 08193 Bellaterra (Barcelona), Spain}

\noindent \hangindent=.5cm $^{35}${Instituci\'{o} Catalana de Recerca i Estudis Avan\c{c}ats, Passeig de Llu\'{\i}s Companys, 23, 08010 Barcelona, Spain}

\noindent \hangindent=.5cm $^{36}${Department of Physics and Astronomy, University of Sussex, Brighton BN1 9QH, U.K}

\noindent \hangindent=.5cm $^{37}${Departamento de F\'{\i}sica, DCI-Campus Le\'{o}n, Universidad de Guanajuato, Loma del Bosque 103, Le\'{o}n, Guanajuato C.~P.~37150, M\'{e}xico}

\noindent \hangindent=.5cm $^{38}${Instituto Avanzado de Cosmolog\'{\i}a A.~C., San Marcos 11 - Atenas 202. Magdalena Contreras. Ciudad de M\'{e}xico C.~P.~10720, M\'{e}xico}

\noindent \hangindent=.5cm $^{39}${Instituto de Astrof\'{i}sica de Andaluc\'{i}a (CSIC), Glorieta de la Astronom\'{i}a, s/n, E-18008 Granada, Spain}

\noindent \hangindent=.5cm $^{40}${Departament de F\'isica, EEBE, Universitat Polit\`ecnica de Catalunya, c/Eduard Maristany 10, 08930 Barcelona, Spain}

\noindent \hangindent=.5cm $^{41}${Department of Physics and Astronomy, Sejong University, 209 Neungdong-ro, Gwangjin-gu, Seoul 05006, Republic of Korea}

\noindent \hangindent=.5cm $^{42}${Abastumani Astrophysical Observatory, Tbilisi, GE-0179, Georgia}

\noindent \hangindent=.5cm $^{43}${Department of Physics, Kansas State University, 116 Cardwell Hall, Manhattan, KS 66506, USA}

\noindent \hangindent=.5cm $^{44}${Faculty of Natural Sciences and Medicine, Ilia State University, 0194 Tbilisi, Georgia}

\noindent \hangindent=.5cm $^{45}${CIEMAT, Avenida Complutense 40, E-28040 Madrid, Spain}

\noindent \hangindent=.5cm $^{46}${National Astronomical Observatories, Chinese Academy of Sciences, A20 Datun Road, Chaoyang District, Beijing, 100101, P.~R.~China}

\bibliographystyle{JHEP}
\bibliography{references}
\end{document}